\documentclass[a4paper,fleqn]{cas-dc}

\usepackage[numbers]{natbib}
\usepackage{color,soul}     
\usepackage[version=4]{mhchem}
\usepackage{upgreek}

\def\tsc#1{\csdef{#1}{\textsc{\lowercase{#1}}\xspace}}
\tsc{WGM}
\tsc{QE}

\newcommand{\HCNp}{\ce{HCN+}}
\newcommand{\HNCp}{\ce{HNC+}}
\newcommand{\HCNHp}{\ce{HCNH+}}
\newcommand{\HCOp}{\ce{HCO+}}

\newcommand{\NN}{\ce{N2}}
\newcommand{\HH}{\ce{H2}}

\newcommand{\eg}{e.\,g.}
\newcommand{\ie}{i.\,e.}

\newcommand{\rcm}{\ensuremath{\text{cm}^{-1}}}

\newcommand{\eV}{\ensuremath{\text{eV}}}

\newcommand{\massU}{\ensuremath{m/z}}

\begin{document}
\let\WriteBookmarks\relax
\def\floatpagepagefraction{1}
\def\textpagefraction{.001}

\shorttitle{}    

\shortauthors{}  

\title [mode = title]{High Resolution Overtone Spectroscopy of \ce{HNC+} and \ce{HCN+}}  



%

\author[1]{Miguel Jiménez-Redondo}
\credit{Conceptualization, Data curation, Investigation, Writing – original draft}

\affiliation[1]{organization={Max Planck Institute for Extraterrestrial Physics},
            addressline={Gießenbachstraße 1}, 
            city={Garching},
            postcode={85748}, 
            country={Germany}}

\author[1]{Chiara Schleif}
\credit{Conceptualization, Data curation, Investigation, Writing – original draft}

\author[2,3]{Julianna Palot{\' a}s}
\credit{Investigation, Writing – original draft}

\affiliation[2]{organization={School of Chemistry, University of Edinburgh},
            addressline={Joseph Black Building, Kings Buildings, David Brewster Road}, 
            city={Edinburgh},
         citysep={}, 
            postcode={EH9 3FJ},
            country={UK}}

\affiliation[3]{organization={Stavropoulos Center for Complex Quantum Matter, Department of Physics and Astronomy, University of Notre Dame}, 
            city={Notre Dame, Indiana},
            citysep={}, 
            postcode={46556},
            country={USA}}

\author[4]{J{\' a}nos Sarka}
\credit{Investigation, Writing – original draft}

\affiliation[4]{organization={I. Physikalisches Institut, Universität zu Köln},
            addressline={Zülpicher Str. 77}, 
            city={Cologne},
            postcode={50937}, 
            country={Germany}}

\author[1]{Hayley Bunn}
\credit{Conceptualization, Data curation, Investigation, Writing – review and editing}

\author[5]{Petr Dohnal}
\credit{Conceptualization, Data curation, Funding acquisition, Writing – review and editing}

\affiliation[5]{organization={Department of Surface and Plasma Science, Faculty of Mathematics and Physics, Charles University},
            addressline={V Hole{\v s}ovi{\v c}k{\' a}ch 2}, 
            city={Prague},
            postcode={18000},
            country={Czech Republic}}

\author[1]{Paola Caselli}
\credit{Conceptualization, Funding acquisition, Writing – review and editing}

\author[1]{Pavol Jusko}
\cormark[1]

\ead{pjusko@mpe.mpg.de}


\credit{Conceptualization, Data curation, Investigation, Project administration, Visualization, Writing – original draft}

\cortext[1]{Corresponding author}



\begin{abstract}
Rotationally resolved spectra of the \HNCp\ and \HCNp\ molecular ions have been recorded in the spectral
range between 6200 and 6800 \rcm\ using a cryogenic ion trap instrument. The rovibrational transitions
were probed using two different action spectroscopy schemes, namely laser-induced reaction (LIR) and
leak-out spectroscopy (LOS). Various vibrational bands of \HNCp\ and \HCNp\ were measured with high
resolution for the first time. For \HNCp, the $\text{X}~^2\Sigma^+~(20^00)-(00^00)$ overtone band was recorded
using LIR, while LOS was used to probe the $\text{X}~^2\Pi~(000)^1-(210)^0\mu$ combination band and
the $\text{X}~^2\Pi~(000)^1-\text{A}~^2\Sigma^+~(10^00)$ vibronic band of \HCNp. Spectroscopic constants, band 
origins and radiative lifetimes for the observed states have been determined. The effective fit for the 
\HCNp\ spectra revealed the presence of strong vibrational couplings leading to perturbations of the
rovibrational levels of the excited states. The two action spectroscopy schemes are compared and their
potential use to explore ion-molecule interactions is discussed. 
\end{abstract}

\begin{highlights}
\item High resolution overtone spectra of \HNCp\ and \HCNp\ ions
\item Action spectroscopy in a cryogenic ion trap
\item Perturbations between the two lowest electronic states of \HCNp   
\item Radiative lifetime measurements for the selected states of \HNCp\ and \HCNp   
\end{highlights}

\begin{keywords}
 \HNCp/\HCNp \sep rovibrational spectroscopy \sep overtone/ combination bands \sep action spectroscopy 
 \sep cryogenic ion trap
\end{keywords}

\maketitle

\section{Introduction}\label{}

\HCNp\ and its lower energy isomer, \HNCp, are important intermediaries in interstellar cyanide chemistry \citep{Loison2014}.
In cold astrophysical environments, the efficient reaction of these
two ions with \HH\ \cite{Dohnal2023} is expected to be the main 
formation mechanism of \HCNHp\ \cite{Fontani2021,Quenard2017}.
In diffuse molecular clouds, where their neutral counterparts, HCN and HNC, are
known to be present \cite{Liszt2001}, \HCNp\ and \HNCp\ can be formed through photoionization by UV photons.
Since a significant
fraction of hydrogen is in
atomic form in these environments,
\HNCp\ and \HCNp\ are primarily lost through dissociative
recombination with electrons \cite{Gans2019}. The absence of the two
ionic isomers from spectroscopic databases \cite{Endres2016} due to
the lack of available high resolution experimental spectra prevents
the identification of these ions in astronomical surveys 
\cite{Quenard2017}, and thus neither \HNCp\ nor \HCNp\ have yet been detected in any galactic or extra-galactic source.

The $\nu_1$ fundamental vibrational transition of \HNCp\ is concurrently reported by \citet{Schmid2025}. Prior to that, there were only a few experimental studies focused on \HNCp\ spectroscopy and no rotationally resolved spectra were available. 
\citet{Gans2019} measured the photoionisation spectra of HNC, where only the $\nu_3$ band of the X $^2\Sigma^+$ electronic ground state of \HNCp\ could be unambiguously assigned with a band origin at 2260 \rcm.
\citet{Forney1992} identified the absorption features of \HNCp\ at 3365.0 \rcm\ as $\nu_1$
and 2195.2 \rcm\ as $\nu_3$ fundamental bands in a neon matrix. The assignment of the $\nu_1$
feature, has been recently disputed by \citet{Schmid2025},
whose high resolution spectra around 3365.0 \rcm\ rather supports the assignment as
the $\nu_1+\nu_2$ combination band of \HCNp.
\textit{Ab initio} calculations for \HNCp\ were performed by \citet{Peterson1990} and later by \citet{Kraemer1992} 
who calculated band origins for the strongest bands of \HNCp\ in the X $^2\Sigma^+$ state for energies of up to 14000 \rcm. 

The rotational spectra for the ground vibrational state of the higher energy isomer, \HCNp, was concurrently to this study measured by \citet{Silva2025} and by \citet{Schmid2025}, who
reported high resolution measurements of the $\nu_1$ fundamental and 
$\nu_1+\nu_2$ combination bands.
Prior to this, the only information on \HCNp\ was based on an infrared spectroscopy study performed in a neon matrix \citep{Forney1992} and on photoelectron spectroscopy of \ce{HCN}  \citep{Baker1968,Frost1973,Fridh1975,Wiedmann1995,Eland1998}.
These experiments covered the three lowest electronic states of \HCNp -- X $^2\Pi$, A $^2\Sigma^+$, and B $^2\Sigma^+$ -- but only the 12 bands with energies of up to 4000 \rcm\ from the ground state obtained by \citet{Wiedmann1995} were measured with sufficient accuracy to resolve individual rotational states. 
\citet{Koppel1979} and later \citet{Taroni2001} used quantum mechanical calculations to assign peaks in the photolectron spectra of \HCNp\ to vibrational bands noting that the spectral assignment is complicated by the strong vibronic coupling between the close lying X $^2\Pi$ and A $^2\Sigma^+$ electronic states.

\citet{Peterson1990} calculated the potential energy surface of the X $^2\Pi$ electronic ground state of \HCNp\ using the configuration interaction (CI) method and reported spectroscopic constants for low lying vibrational levels. 
The coupled cluster study by \citet{Botschwina1997} covered both the X $^2\Pi$ and A $^2\Sigma^+$ electronic states but only the linear geometry was considered in the calculations. 
Higher lying electronic states and non-linear geometry of the \HCNp\ molecular ion were investigated in \textit{ab initio} studies by \citet{Hirst2005} and by \citet{Anusuri2016}.
Potential energy curves for the two lowest electronic states of \HCNp\ and \HNCp\ were calculated using the CASSCF method by \citet{Talbi2000} in connection with their theoretical investigation of the dissociative recombination of these ions with electrons.

\begin{figure}[h]
\centering
  \includegraphics[width=0.98\linewidth]{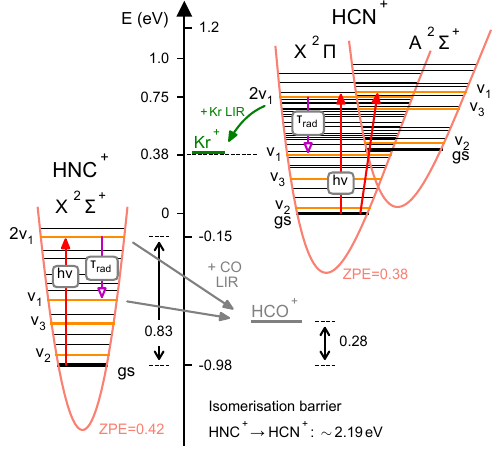}
  \caption{Schematic overview of the lowest vibronic states of \HCNp\ and \HNCp. The vibrational energies and the zero point energies (ZPE) were taken from refs. \citep{Kraemer1992,Taroni2001}. Only a subset of vibrational levels assigned by \citet{Taroni2001} are plotted. The laser-induced reaction probing schemes involving proton transfer to CO for \HNCp\ and charge transfer in reaction with Kr for \HCNp\ are denoted by arrows. Potential energy minima and curve crossing for the X~$^2\Pi$ and A~$^2\Sigma^+$ states of \HCNp\ are not to scale to improve readability.     
  }\label{f_levels}
\end{figure}

The X $^2\Pi$ and A $^2\Sigma^+$ electronic states of \HCNp\ as well as the X $^2\Sigma^+$ state of \HNCp\ are strongly bound with dissociation energies of 6.4~\eV, 5.4~\eV\ and 6.7~\eV, respectively \citep{Talbi2000}. The energy difference between the ground states of \HNCp\ and \HCNp\ is 0.98~\eV \citep{Hansel1998} while the  barrier for isomerisation from \HNCp\ to \HCNp\ is 2.188~eV \citep{Taroni2001}. 
Both isomers present a linear geometry in their ground states.
A schematic energy level diagram for \HCNp\ and \HNCp\ is shown in Figure \ref{f_levels}.

The previous lack of high resolution spectra of \HNCp\ and \HCNp\ is especially pronounced in comparison with neutral \ce{HCN} whose overtone transitions around $1.5\;\upmu\text{m}$ were
first measured more than 70 years ago \cite{Wiggins1954} and are routinely used for calibration purposes \citep{Sasada1990,Howard2024}.
This study focuses exactly on this wavelength range and reports experimental, rotationally resolved vibrational spectra of the
X $^2\Sigma^+~(00^00)-(20^00)$ overtone of \HNCp\ and
of the X $^2\Pi~(000)^1-(210)^0\mu$ combination band and the X $^2\Pi~(000)^1-\text{A}
~^2\Sigma^+~(10^00)$ vibronic band of \HCNp.

\section{Methods}

\subsection{Spectroscopic notation}

Here and in the following text, the transitions between rovibrational levels of the \HCNp\ and \HNCp\ ions 
are denoted as $^{\Delta N}\Delta J_{F'_nF''_np''}(J'')$, where $\Delta J$ and $\Delta N$ describe the change 
of the total angular momentum and of the total angular momentum without electron spin, respectively, 
$F_n$ denotes the spin-orbit components and $p$ represents the parity, when applicable. 
The prime and double prime distinguishes quantum numbers belonging to the upper and lower level, respectively.
The vibrational energy levels are denoted by the corresponding vibrational quantum numbers 
$(\nu_1,\nu_2,\nu_3)^Ki$ for the $^2\Pi$ state of \HCNp\ and $(\nu_1,\nu_2^l,\nu_3)$ for
the $^2\Sigma^+$ electronic states of both \HCNp\ and \HNCp.
Here, $K=\Lambda + l$ and $\Lambda$ and $l$ are the electronic angular momentum and vibrational angular momentum 
quantum numbers, respectively. The $i$ distinguishes between lower ($\mu$) or upper 
($\kappa$) levels arising from the Renner-Teller interaction in the $^2\Pi$ state.  

\subsection{Experimental}

The experiments have been conducted in the 22 pole Cold CAS Ion Trap (CCIT) setup \cite{Jusko2024}. The \HCNp\ ions are produced using electron bombardment of \ce{HCN} precursor
gas in a Storage Ion Source (SIS) and then mass selected using a quadrupole mass filter prior to
injection into the 22 pole trap, where an intense He pulse is used to trap and cool them. In order to
maintain sufficient vapor pressure of the introduced neutral gases and to populate
multiple rotational states in the ground vibrational state of the stored ions, the trap
temperature is held constant around either 90 or $125\;\text{K}$, depending on the neutral gas, using a resistive heater .
The ions leaving the trap are then mass selected by a second quadrupole mass filter and
counted using a Daly type detector.
In order to study the lower energy isomer \HNCp, a small amount of \ce{CO2} was added to
the He pulse to isomerize the \HCNp\ ions coming from the SIS,
as described in \citet{Dohnal2023}.

An Agilent 8164B option 200 light system with line width well below $1\;\text{MHz}$
and output power $1-8\;\text{mW}$ was used as a light source,
the wavelength has been calibrated using an EXFO WA 1650 wavemeter, which has an absolute accuracy better than 0.3 ppm.
Two action spectroscopy schemes are used to record the rovibrational transitions
throughout this work: 1.) a laser-induced reaction (LIR) scheme
mainly for \HNCp, where activation of an endothermic ion-molecule reaction upon excitation 
into a vibrational state is used (for details see ref. \cite{JimenezRedondo2024}) 
and 2.) a leak-out spectroscopy (LOS) scheme mainly for \HCNp, where the vibration-translation 
(V-T) transfer upon collision with a neutral buffer gas is applied (for details see ref. \cite{Schmid2022}).
The two schemes are compared in section~\ref{sec_lir_los}.

\subsection{Band origin predictions}
\label{sec:Computational}

The experimental observation of weak (overtones/ combination bands) modes can be very tricky, especially in case of action spectroscopy, where it is unclear if the selected action scheme is viable. 
This is even more prominent in a case like this, where no previous data on the frequency of such bands nor the action scheme is available. Therefore, in order to guide frequency measurements,
anharmonic vibrational frequencies
were calculated for the overtone of $\nu_1$ in both \HCNp\ and \HNCp\ ions
using second-order vibrational perturbation theory (VPT2).
Coupled-cluster theory has been utilized including single and double excitations
and a perturbative treatment of the triple excitations, CCSD(T),
using both RHF or UHF reference functions
with augmented correlation-consistent basis sets up to aug-cc-pV5Z.
All the calculations in this work were performed using the CFOUR program suite \cite{cfour,08HaMeGaAu,20MaChHaLi}.
For \HCNp, the full characterization of the rovibrational spectrum would require
the consideration of the Renner-Teller effect and the spin-orbit coupling.
Nevertheless, describing only the $\nu_1$ overtone accurately can be done more easily by considering only a single electronic state, which has been performed here and was sufficient to guide our experimental search.

\section{Results and Discussion}

\subsection{HNC$^+$ spectroscopy}

The rovibrational spectrum of the X $^2\Sigma^+~(00^00)-(20^00)$ overtone of \HNCp\ was recorded using the LIR scheme
\begin{equation}
\ce{HNC+} + \ce{CO} \to \ce{HCO+} + \ce{CN} ~~~~~\Delta H_0=0.28\;\eV,
\label{eq:lir_hnc_overtone}
\end{equation}
where the reaction enthalpy at 0 K was taken from Active Thermochemical Tables (ATcT) \citep{Ruscic2005}. 
The transition involving unresolved doublet $^\mathrm{r}\mathrm{R}_{11}(4.5)$/$^\mathrm{r}\mathrm{R}_{22}(3.5)$ was additionally probed using leak-out spectroscopy (LOS) \citep{Schmid2022} 
with \ce{N2} as collisional partner for the transfer of vibrational excitation 
(see section~\ref{sec_lir_los}). 
In this section only the LIR spectroscopic data is considered, as the signal-to-noise ratio in the 
LOS experiments was much lower.

\begin{figure}[h]
\centering
  \includegraphics[width=0.98\linewidth]{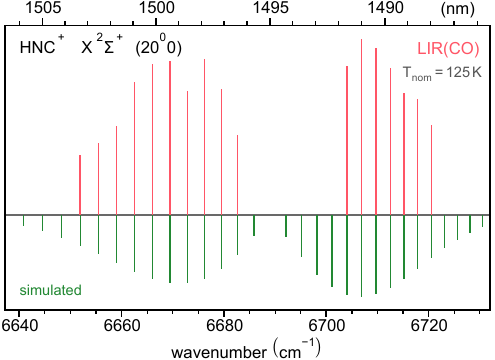}
  \caption{Comparison of the measured central wavenumbers for transitions in the $(20^00)$ band of \HNCp\ (upper trace) and of the spectrum generated by PGOPHER \citep{Western2017} using the spectroscopic constants listed in Table \ref{t_constants_HNCp} (lower trace). The displayed experimental line intensities were normalized to the number of primary ions and power of the laser.  
  }\label{f_hnc}
\end{figure}

Absorption line profiles were measured by recording the number of trapped \HCOp\ ions produced in the LIR (\ref{eq:lir_hnc_overtone})
as a function of the laser frequency. 
A fit to a Doppler function was then used to
obtain accurate transition wavenumbers, which are listed in Table S1 in the SI and plotted 
in Figure~\ref{f_hnc}.
The energy levels for the upper and lower states involved in the measured overtone transitions of \HNCp\ were described by a standard Hamiltonian for a linear molecule \cite{JimenezRedondo2024}:
\begin{equation}
\label{eq_Hamiltonian}
H = T_\nu+B_\nu N (N+1)-D_\nu [N(N+1)]^2,
\end{equation}
where $\nu$ denotes the vibrational quantum numbers, $B_\nu$ and $D_\nu$ are rotational and distortion constants, respectively, $N$ is the total angular momentum without electronic spin and $T_\nu$ is the vibrational energy term set to zero for the vibrational ground state. $N = J-0.5$ for $e$ states ($F_1$) and $N=J+0.5$ for $f$ states ($F_2$). 
The spin rotational interaction due to $^2\Sigma$ character of the ground electronic state is not considered in the data analysis as it was not visible in the measured spectrum. The line positions listed in Table S1 in the SI correspond to the mean wavenumber for each unresolved doublet.

The data was fitted using the rotational Hamiltonian (\ref{eq_Hamiltonian}) implemented in PGOPHER\citep{Western2017}. The obtained spectroscopic constants are shown in Table~\ref{t_constants_HNCp} and compared to values from theoretical \citep{Peterson1990,Kraemer1992} and experimental \citep{Schmid2025} studies. The agreement between present spectroscopic constants for the ground state and those determined by \citet{Schmid2025} from the spectra of the $^2\Sigma^+~(00^00)-(10^00)$ fundamental band is within error.  
Note that \citet{Kraemer1992} did not publish rotational constants for \HNCp, only energies for rotational states up to $N=6$ were reported. 
The corresponding values of $B_{00^00}$ and $D_{00^00}$ in Table \ref{t_constants_HNCp} were obtained by fitting the rotational levels from ref. \citep{Kraemer1992} with equation (\ref{eq_Hamiltonian}).
\begin{table*}[ht]
    \caption[]{Spectroscopic constants for the ground and the $^2\Sigma^+~(20^00)$ vibrational state of the electronic ground state of \HNCp\ determined from the measured data, along with the calculated values derived in this work at VPT2 using CCSD(T) with RHF and UHF reference functions, the calculations by \citet{Peterson1990}, by \citet{Kraemer1992}, and recent experimental data by \citet{Schmid2025}.} 
    \label{t_constants_HNCp} 
    \begin{tabular}{ c c c c c c c}
        \toprule
              & This work     & This work & This work
              & Ref. \citep{Peterson1990} & Ref. \citep{Kraemer1992} & Ref. \citep{Schmid2025}\\
              & (experiment)     & RHF-CCSD(T) & UHF-CCSD(T)
              & (theory) & (theory) & (experiment)\\
    \midrule
      $B_{00^00}$ & 1.571687(56)&  1.566 & 1.574 & 1.5718(23) & 1.549(10) & 1.57169(4)\\
      $D_{00^00}$ & 3.16(33)$\times10^{-6}$ & 3.06$\times10^{-6}$ & 2.96$\times10^{-6}$ & & 3.1$\times10^{-6}$ & 3.10(26)$\times10^{-6}$\\
      $T_{20^00}$ & 6689.01950(81) & 6670.5 & 6691.1 &  6800 & 6685.49 & \\
      $B_{20^00}$ & 1.548292(49) & 1.542 & 1.551 &  1.5449 & &\\
      $D_{20^00}$ & 2.82(29)$\times10^{-6}$& & & & & \\
        \noalign{\smallskip}
        \bottomrule
    \end{tabular}
    \vskip 0.2em
    {\footnotesize{\emph{Note:}
    All values in units of \rcm. Numbers in parentheses are statistical errors of the fit in units of the last quoted digit.
    }}
\end{table*}

The measured value of the X $^2\Sigma~(00^00)-(20^00)$ band origin $T_{20^00} = 6689.01950(81)$~\rcm\ is very close to the value of 6685.49~\rcm\ predicted by \citet{Kraemer1992} and to the calculations done for this work (see section \ref{sec:Computational}) at FC-CCSD(T)/aug-cc-pV5Z using the UHF reference functions, resulting in a value of 6691.1 \rcm. The use of RHF reference functions obtained a prediction that was with 6670.5~\rcm\ slightly further off, and the band origin calculated by \citet{Peterson1990} turned out to be more than 100~\rcm\ lower than the determined position. This is not surprising as the latter theory considered only a linear geometry for the \HNCp\ ion. For the rotational constants, the calculation
by \citet{Peterson1990} for the ground state presents the best
agreement with the present results, and their predicted value for
$B_{20^00}$ also compares favorably to the experimental one.

\subsection{HCN$^+$ spectroscopy}

A LIR technique based on the charge transfer reaction  
\begin{equation}
\ce{HCN+} + \ce{Kr} \to \ce{Kr+} + \ce{HCN} ~~~~~\Delta H_0=0.39\;\eV,
\label{eq:lir_hcn_lir}
\end{equation}
where $\Delta H_0$ is the reaction enthalpy at 0 K, taken from ATcT \citep{Ruscic2005},
was initially used to detect six transitions of \HCNp\ in the spectral region around 6330~\rcm.
As \ce{Kr+} ions further reacted with impurities in the krypton gas (mainly \ce{O2}), negatively affecting 
the signal to noise ratio, a leak-out spectroscopy (LOS) scheme with \ce{N2} as neutral buffer gas was then employed to measure the
majority of the \HCNp\ transitions reported here. 
 
The rovibrational energy levels of the X~$^2\Pi$ electronic ground state were described by the following Hamiltonian:
\begin{equation}
\hat{H} =\hat{H}_\mathrm{VIB} + \hat{H}_\mathrm{ROT} + \hat{H}_\mathrm{SO} + \hat{H}_\mathrm{SR} + \hat{H}_\mathrm{\Lambda},
\label{eq:Hamiltonian_HCN+X_all}
\end{equation}
where $\hat{H}_\mathrm{VIB} = T\hat{1}$ with $T$ being the vibrational term energy, $\hat{1}$ the identity operator, and
\begin{equation}
\hat{H}_\mathrm{ROT} = B\hat{\mathrm{N}}^2-D\hat{\mathrm{N}}^4+H\hat{\mathrm{N}}^6,
\label{eq:Hamiltonian_Hrot}
\end{equation}
\begin{equation}
\hat{H}_\mathrm{SO} = A\hat{L}_\mathrm{z}\hat{S}_\mathrm{z},
\label{eq:Hamiltonian_HSO}
\end{equation}
\begin{equation}
\hat{H}_\mathrm{SR} = \gamma\hat{\mathrm{N}}\cdot\hat{\mathrm{S}}+\frac{1}{2}\gamma_D \left[\hat{\mathrm{N}}\cdot\hat{\mathrm{S}},\hat{\mathrm{N}}^2 \right]_+,
\label{eq:Hamiltonian_HSR}
\end{equation}
\begin{equation}
\begin{split}
\hat{H}_\mathrm{\Lambda} = &-\frac{1}{2}p\left(\hat{N}_+\hat{S}_+e^{-2i\phi}+\hat{N}_-\hat{S}_-e^{+2i\phi} \right) \\
&+\frac{1}{2}q\left( \hat{N}_+^2e^{-2i\phi}+\hat{N}_-^2e^{+2i\phi}\right) \\
&-\frac{1}{4}\left[ \hat{N}_+\hat{S}_+e^{-2i\phi}+\hat{N}_-\hat{S}_-e^{+2i\phi}, p_D\hat{\mathrm{N}}^2 \right]_+ \\
&+\frac{1}{4}\left[ \hat{N}_+^2e^{-2i\phi} + \hat{N}_-^2e^{+2i\phi}, q_D\hat{\mathrm{N}}^2\right]_+,
\label{eq:Hamiltonian_Lambda}
\end{split}
\end{equation}
 to a Hund's case (a) basis as implemented in PGOPHER\citep{Western2017}. $\hat{H}_\mathrm{ROT}$ describes the rotational energy term where $B$ denotes the rotational constant, and $D$ and $H$ the quartic and sextic centrifugal distortion constants, respectively. $\hat{H}_\mathrm{SO}$ represents the spin-orbit interaction including the spin-orbit coupling constant $A$, while $\hat{H}_\mathrm{SR}$ describes the spin-rotation interaction with $\gamma$ denoting the spin-rotation coupling constant and $\gamma_D$ its centrifugal distortion. $\hat{H}_\mathrm{\Lambda}$ represents the $\Lambda$-type doubling containing the two $\Lambda$ doubling constants $p$ and $q$ as well as their centrifugal distortion parameters $p_D$ and $q_D$. The terms $e^{\pm2i\phi}$ assure that the operators of the $\Lambda$-type doubling connect only the two halves of a $\Pi$ state. The square brackets denote the anti-commutator
\begin{equation}
\left[ \hat{O},\hat{Q} \right]_+ = \hat{O}\hat{Q}+\hat{Q}\hat{O}.
\end{equation}

The Hamiltonian for the A~$^2\Sigma^+$ state of \HCNp\ is similar to that of the ground electronic state (\ref{eq:Hamiltonian_HCN+X_all}) but without Spin-Orbit and $\Lambda$ doubling terms:
\begin{equation}
\hat{H} = \hat{H}_\mathrm{VIB} + \hat{H}_\mathrm{ROT} + \hat{H}_\mathrm{SR}.
\label{eq:Hamiltonian_HCN+A_all}
\end{equation}

\begin{figure*}[h]
\centering
  \includegraphics[]{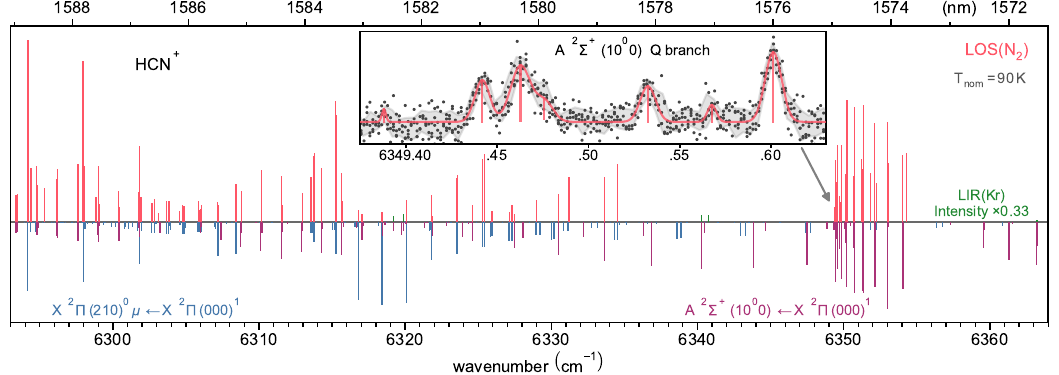}
  \caption{Comparison of the measured central wavenumbers for transitions of the bands X~$^2\Pi$
  $(210)^0\mu$ and A~$^2\Sigma^+~(10^00)$ of \HCNp\ (upper trace) and of the spectrum 
  simulated by PGOPHER \citep{Western2017} using the spectroscopic constants in Table \ref{t_constants_HCNp} (lower trace). 
  The displayed experimental intensities were normalized to the number of primary ions and power of the laser. 
  LOS intensity is scaled 3~times to match the LIR intensity on the same transitions. 
  The inset is showing a section of the Q branch rovibronic transitions, the red line
  depicts the fitted function determining the transition centers (sticks).
  Note that not the entire area has been continuously scanned (see text).
  }\label{f_hcn}
\end{figure*}

The spectral region between 6292 and 6364~\rcm\ was investigated using
either the LIR or LOS action schemes. Due to the overall lack of detected absorption features in the regions surrounding them, the wavenumber ranges 6342--6348, 6355--6357, and 6361--6363~\rcm were not scanned. The analysis
of the absorption line profiles is analogous to
the \HNCp\ case. All the measured transitions of the \HCNp\ ion are listed in Table S2 in the SI and plotted in Figure~\ref{f_hcn} as a stick spectrum indicating line positions and relative intensities. 

The majority of the observed transitions are attributed to two bands of the \HCNp\ ion. The progression of the band centered at 6330.5924(36)~\rcm\ is characteristic for a $^2\Pi-^2\Sigma$ system with a strong Q$_{11}$ branch between 6349 \rcm\ and 6355 \rcm. 
Due to vibronic interactions between the lowest electronic states of \HCNp, X~$^2\Pi~(1 ^2A', 1 ^2A'')$ and A~$^2\Sigma^+~(2 ^2A')$, the electronic $\Lambda$ and vibrational $l$ angular momentum quantum numbers are no longer good quantum numbers \citep{Taroni2001}, however their sum $K = \Lambda + l$ is still conserved.
Levels with the same $K$ can in principle
mix, leading to the vibrational quantum numbers $\nu_1$, $\nu_2$ and $\nu_3$ losing significance. 
Thus the assignment of particular energy levels to vibrational quantum numbers is not straightforward
and in the case of strong vibronic coupling corresponds rather to the most similar energy levels of the unperturbed model system \citep{Taroni2001}.
\citet{Taroni2001} assigned the calculated vibrational energy level at 6345~\rcm, with respect to ground state, as X$(210)^0\mu$ and A$(10^00)$.
Based on the $^2\Pi-^2\Sigma$ character of the observed band, the above mentioned calculations of \citet{Taroni2001}, and as there are no other A state levels predicted in the vicinity, 
this band is assigned as X $^2\Pi~(000)^1-\mathrm{A}^2\Sigma^+~(10^00)$.

The spectroscopic constants obtained for the A $^2\Sigma^+~(10^00)$ state by including 51 of the measured 113 transitions into the fit are shown in Table \ref{t_constants_HCNp}. 
The constants for the ground state were fixed at a value taken from refs. \citep{Schmid2025,Silva2025}.
Only $T$, $B$, $D$ and $\gamma$ were determined for the A $^2\Sigma^+(10^00)$ state, resulting in an average fitting error of 0.0098~\rcm.

Adding higher order corrections to the fitted Hamiltonian, in particular the centrifugal distortion parameter $\gamma_D$ of the spin-rotation coupling constant $\gamma$, 
led to a slight decrease of the average fitting error, but simultaneously to an increase of the relative error of $D$ to above $100\%$. 
$D$ was, consequently, set to zero for this fitting approach, resulting in an average error of 0.0089 \rcm.

The other band with origin at 6284.9187(19) \rcm\ is assigned as X $^2\Pi~(000)^1-(210)^0\mu$ based on its observed $^2\Pi-^2\Pi$ character and the calculations of \citet{Taroni2001}. 
The spectroscopic constants fitted to 50 of the measured 113 transitions are listed 
in Table~\ref{t_constants_HCNp}. 
Two additional transitions, one as part of an unresolved doublet, were assigned reliably, but had to be removed from the effective fit as they were observed to remain at high residuals. The two transitions are marked with an asterisk (*) in Table S2 in the SI.

The determined rotational constant $B_{210}$ = 1.346376(71)\;\rcm\ is slightly lower than that of the ground state $B_{000} =$ $1.35275(1)\;\rcm$ \citep{Schmid2025, Silva2025} as expected for vibrational levels of the same electronic state.
The low value determined for the spin-orbit coupling constant $A$ supports the spectroscopic assignment as the strong coupling between the $^2\Pi$ and $^2\Sigma^+$ states is expected to partially inhibit the spin-orbit interaction, 
which is further quenched for levels with $K<\nu_2+1$ \citep{Wiedmann1995}.  
It has to be considered that for pure $^2\Pi - ^2\Pi$ systems, the transitions with $\Delta\Omega\neq0$ are dipole-forbidden \citep{Augustovicova2024}. 
The observation of these transitions in the present study is, therefore, a further indication of coupling between the X~$^2\Pi$ and A~$^2\Sigma^+$ states.

The inclusion of higher order correction terms (containing the centrifugal distortion parameters $p_D, q_D$ and $\gamma_D$) to the Hamiltonian led to an improvement of the average fitting error from 0.00548 \rcm\ to 0.0012 \rcm\ 
and enabled the inclusion of the two previously excluded transitions (marked with an asterisk (*) in Table S2). 

The large fitting error that remains for both bands when excluding higher order correction terms is substantially higher than the accuracy of the present experimental setup. This indicates a strong interaction between the two identified states and furthermore extensive vibronic coupling with other nearby, unobserved states, as the effective Hamiltonian utilized for the fits cannot reproduce all the observed line positions. 
Unfortunately, the attempt of implementing corresponding perturbation terms into the simulation did not result in a meaningful improvement of the fitting error.
The effective fit reported in this study does, therefore, only consider unperturbed transitions.\\

Especially in the case of $(210)^0\mu$, part of the perturbations seem to be efficiently ``absorbed'' by the higher order corrections thus leading to the more significant improvement of the fit.
In both cases, the physical interpretation of the derived spectroscopic parameters $p_D, q_D$ and $\gamma_D$ should be taken with caution. Furthermore, considering the spectroscopic resolution of the measurement procedure, it is unlikely to reliably fit such higher order constants to the experimental data reported in this work. 
The extended sets of spectroscopic constants were, consequently, not utilised in the final fits and are only reported in Table S3 in the SI for comparability.

\begin{table}[ht]
    \caption[]{Spectroscopic constants for the $\mathrm{X}~(210)^0\mu$ and $\mathrm{A}~^2\Sigma^+~(10^00)$ states of \HCNp\ fitted to 50 and 51 of the observed transitions, respectively, using the Hamiltonian (\ref{eq:Hamiltonian_HCN+X_all}) and ground state constants from ref. \citep{Schmid2025}.}\label{t_constants_HCNp}
    \begin{tabular}{ c c c }
        \toprule
              & $\mathrm{X}~(210)^0\mu$     & $\mathrm{A}~^2\Sigma^+~(10^00)$ \\
        \midrule
        $T$ & 6284.9187(19)& 6330.5924(36)  \\
        $B$ & 1.346376(71)  & 1.38624(13) \\
        $D$ & 2.92(49)$\times10^{-6}$ & 9.4(82)$\times10^{-7}$  \\
        $A$ & 0.342(11) &  \\
        $\gamma$ & 0.02039(33) & 0.01305(40)  \\
        $p$ & 0.03685(51) &  \\
        $q$ & 0.002387(27) &  \\
        \noalign{\smallskip}
        \bottomrule
    \end{tabular}
    \vskip 0.2em
    {\footnotesize{\emph{Note:}
    All values in units of \rcm. Numbers in parentheses are statistical errors of the fit in units of the last quoted digit.
    }}
\end{table}

Of the 113 measured transitions, 51 were assigned to the X $^2\Pi~(000)^1-(210)^0\mu$ band and 
51 to the X $^2\Pi~(000)^1-\mathrm{A}^2\Sigma^+~(10^00)$ band of \HCNp. 
The presence of unaccounted vibronic and rovibrational couplings makes the use of quantum numbers associated with an effective Hamiltonian challenging. 
This is especially apparent in the case of the A~$^2\Sigma^+~(10^00)$ state where the average error of the fit given 
by the spectroscopic constants from Table \ref{t_constants_HCNp} was 0.0098 \rcm. 
The largest difference between the measured and calculated line positions was observed for the lowest $J'$ states. As an example, the simulated transitions closest to that observed at 6354.01791(32) \rcm\ are $^\mathrm{q}\mathrm{Q}_{11}(1.5)$ at 6354.076 \rcm\  and $^\mathrm{q}\mathrm{P}_{21}(1.5)$ at 6354.056 \rcm, resulting in a residual of either 0.058 or 0.038 \rcm. 
Furthermore, it can be seen that for a few doublets with low $J'$, the simulated intensities seem to be "switched" compared to the measured transitions. Although intensities measured with these experimental techniques are often unreliable due to additional factors relating to the collision process, this in turn could suggest that in reality, the transitions need to be assigned conversely.
These observations indicate that energy levels of the upper state with low $J'$ quantum numbers are significantly perturbed with respect to those predicted by the effective Hamiltonian for the $^2\Sigma^+$ state without considering perturbations from nearby bands. 

The 2$\nu_1$ overtone band of \HCNp\ was also considered as a target 
for the present spectroscopic study.
The calculation from \citet{Peterson1990}, which, 
as mentioned before for \HNCp, neglected 
the bend-stretch interaction,
predicted the band origin at 6090~\rcm.
Our calculation at FC-CCSD(T)/aug-cc-pV5Z
with RHF reference places it at 6003.3~\rcm, while using UHF reference shifts it to 5989.0~\rcm.
This unfortunately falls outside of the capability of the laser system used in this experimental setup. 

\subsection{Radiative lifetime determination}

The radiative lifetime of the upper level of the probed transition can be estimated 
from the dependence of the number of LIR product ions
(\ie, the line intensity) on the number density of the LIR reactant [M]. 
In the case where the levels with energies below that of the upper state of the transition do 
not react with the LIR reactant gas, a simple relation can be obtained \citep{JimenezRedondo2024}:
\begin{equation}
    \frac{N_{\mathrm{LIR}}}{N_\mathrm{P}}=\frac{r_1k_1[\ce{M}]}{1/\tau+k_1[\ce{M}]}.
    \label{eq_tau_determination}
\end{equation}
where $r_1$ is the rate of excitation from the lower state to the upper state, $k_1$ is the rate 
coefficient for the given laser induced reaction, assumed to have a value close to the collisional (Langevin) coefficient,
$N_\mathrm{P}$ is the number of primary ions and $\tau$ is the lifetime of the upper state.

If the LIR is exothermic for some of the energy levels accessible from the upper state by
radiative decay, a simple analysis by equation (\ref{eq_tau_determination}) is no longer sufficient for a single
state lifetime determination. The 
$\tau$ derived from equation (\ref{eq_tau_determination}) then represents the ``effective'' lifetime for the 
radiative cascade starting in the upper level of a given transition and ending in some lower state with 
insufficient energy for the probing reaction. The obtained $\tau$ is thus only an upper estimate of the 
real lifetime.

The endothermicities of reactions (\ref{eq:lir_hnc_overtone}) and (\ref{eq:lir_hcn_lir})  used in the present spectroscopic
study for \HNCp\ and \HCNp\ are relatively low (\ie, the fundamental mode can
still lead to the LIR process, see Figure~\ref{f_levels}) leading to the situation described in the previous paragraph.
The effective lifetime of the X $^2\Sigma^+~(20^00)$ vibrational state of \HNCp\ was determined using equation
(\ref{eq_tau_determination}) from the dependence of the ratio of $N_{\ce{HCO+}}/N_{\HNCp}$ on [CO]. 
The unresolved $^\mathrm{r}\mathrm{R}_{11}(4.5)$/$^\mathrm{r}\mathrm{R}_{22}(3.5)$ transition was used and $\tau = 0.401 \pm 0.085$~ms was
obtained at $T = 125\;\text{K}$.
For comparison, the transition dipole moment (matrix element $\mu_\mathrm{z}$ between respective vibrational 
basis functions) for the $\nu_1$ fundamental of \HNCp, calculated by \citet{Kraemer1992} as 
$\mu_\mathrm{z} = -0.239$ Debye, corresponds to a radiative lifetime for the X $^2\Sigma^+~(10^00)$ state 
of 1.4~ms. To approximately compare to the present data, the 1/$\nu$ scaling law, exactly valid only in harmonic approximation, can be applied \citep{Mauclaire1995}, i.e., the $(10^00)$ state should have a lifetime about two times longer than the $(20^00)$ state. The resulting estimated lifetime is higher than the measured one but still reasonably close.   

In the case of \HCNp, the endothermicity of reaction (\ref{eq:lir_hcn_lir}) is extremely close (albeit 
slightly higher) to the energy of the $\nu_1$ vibration reported by \citet{Forney1992}. 
The value of the effective lifetime of the \HCNp\ vibrational states reported here is, therefore, most 
probably close to the real radiative lifetime of these states. 
The $\tau = 0.379 \pm 0.078~\mathrm{ms}$ for the X~$(210)^0\mu$ state was determined at $125\;\text{K}$
from the transition centered at 6320.07994(15)~\rcm\ identified as an unresolved 
doublet $^\mathrm{r}\mathrm{R}_{1e/f}(5.5)$.  
The effective lifetime of the A~$^2\Sigma^+~(10^00)$ vibrational state was obtained using the
transition $^\mathrm{q}\mathrm{Q}_{1}(8.5)$ at 6349.6012 \rcm\ to be $\tau = 0.52 \pm 0.25~\mathrm{ms}$
and a similar value of $\tau = 0.47 \pm 0.22~\mathrm{ms}$ was determined 
using the transition $^\mathrm{q}\mathrm{Q}_{1}(5.5)$ at 6350.7346 \rcm.
See Section S2 in the SI for details.

\subsection{Action scheme comparison LIR/LOS}
\label{sec_lir_los}

The effectiveness of both LIR and LOS techniques is connected to the collisional interaction of the 
vibrationally excited ions with the corresponding neutrals. 
The comparison between ion signal intensities arising from LIR reaction products, or from ions leaked out 
of the trap during the LOS, provides useful indirect information about the underlying processes.
On one hand, energetically favored ion-molecule reactions, such as charge or proton transfer, regularly
applied in LIR schemes, often proceed at collisional reaction rates. 
On the other hand, the efficiency of the LOS scheme is inherently linked to the rate of collisional 
quenching of the ions' vibrationally excited states, \ie, to the vibration-translation (V-T) energy transfer.
The signal intensity from these two techniques is, therefore, compared while maintaining all the
remaining parameters equal, in order to constrain the ratio between the two processes 
(collisional rate vs. V-T transfer).  

\begin{figure}[h]
\centering
  \includegraphics[width=0.98\linewidth]{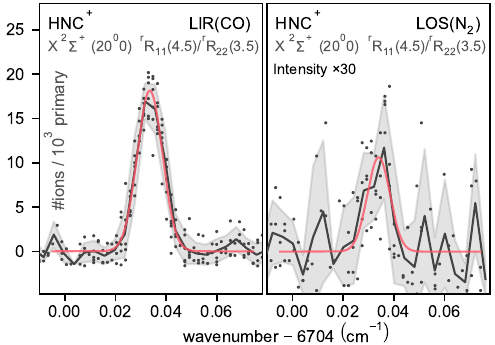}
  \caption{Absorption line profiles for the unresolved transitions $^\mathrm{r}\mathrm{R}_{11}(4.5)$/$^\mathrm{r}\mathrm{R}_{22}(3.5)$ of the X $^2\Sigma^+~(00^00)-(20^00)$ band of \HNCp\ obtained by LIR and LOS techniques. In both cases, the constant background signal was subtracted to enable comparison of line intensities. 
  The data is normalized to the number of primary ions. 
  All the remaining parameters (laser power, irradiation period, temperature, rf amplitude, etc.) were kept constant. 
  }\label{f_lirlos}
\end{figure}

The comparison of the transitions $^\mathrm{r}\mathrm{R}_{11}(4.5)$/$^\mathrm{r}\mathrm{R}_{22}(3.5)$ 
 of the X $^2\Sigma^+~(00^00)-(20^00)$ band of \HNCp\ in Figure \ref{f_lirlos} shows a 
factor of 50 difference between the ion signals measured by LIR and LOS techniques,
while keeping all other parameters the same.
This indicates that the rate coefficient for collisional quenching of the 
X $^2\Sigma^+~(20^00)$ state of \HNCp\ by \ce{N2} is substantially lower 
than the corresponding Langevin reaction rate coefficient for reaction (\ref{eq:lir_hnc_overtone}).
For comparison, there was only a factor of three difference between the LIR and LOS ion signals 
in the present spectroscopic study of \HCNp\ (within the same electronic state).

The ion signal attributed to the action scheme during LIR or LOS experiments is affected by 
a variety of factors, most obvious being the rf and dc potentials used for ion trapping and extraction 
from the trap (note that LIR uses only extraction of ions at the end of the trapping cycle, while
LOS uses the ``leak-out'' of ions during the whole irradiation period). 
Although, to our knowledge, the approach to estimate the ratio between the collisional rate and 
the V-T transfer presented in Section~\ref{sec_lir_los} has not been explored so far,
the \ce{[CHN]+} isomeric system gives a unique opportunity to develop these kind of methods,
because of access to two completely different systems (isomers), 
which share the same ion mass $27\;\massU$, \ie, the same behavior inside the 22 pole rf trap.
Moreover, in case of \HNCp, even the LIR product ion \HCOp\ is only $2\;\massU$ heavier. 
In this way, many of the effects, which could easily be miss-attributed to the change of 
the effective/ extraction potentials or to the ion detector due to the ion mass/ charge ratio change simply even out.
We believe our result, that estimates that the V-T transfer of \HNCp\ to \NN\ is
approx. 10 times less efficient than for \HCNp\ to \NN\ is warranted.
The most probable explanation for this behavior is the high density of states of \HCNp, \ie, higher probability
of better vibration-vibration coupling during the collision with \ce{N2}.
This method, capable of assessing various ion-molecule interaction properties, shall certainly
be explored further.

\section{Conclusion}

We provide the first data for the overtone spectra of \HNCp\ and \HCNp\ ions. Together
with the simultaneous publications for the fundamental vibrations \cite{Schmid2025} and the
pure rotational spectrum of \HCNp\ \cite{Silva2025},
these are the first high-resolution spectra available for perhaps the simplest
isomer system of \ce{[CHN]+}, only possible thanks to the latest developments in action spectroscopy and
cryogenic ion trap technology. 
Moreover, the action spectroscopy techniques open new possibilities of ion-molecule 
interaction studies, such as the radiative lifetimes in excited vibrational states and/ or
for the first time also to constrain the efficiency of the vibration-translation (V-T) transfer. 

The two lowest electronic states of \HCNp\ represent a very peculiar spectroscopic system. On the one hand, the lower electronic state is doubly degenerate in linear geometry, resulting in Renner-Teller splitting when the bending mode is excited.
On the other hand, the presence of the low lying A~$^2\Sigma^+$ electronic state leads to strong vibronic coupling between these two states \citep{Taroni2001}.
As a relatively simple molecule for computational purposes, it is a prospective target for potential testing of 
high precision quantum mechanical calculations.

All the high resolution data collected here and in \citet{Schmid2025, Silva2025}, 
together with the derived spectroscopic constants, 
should finally enable the detection of the \HNCp\ and \HCNp\ in space, either by
using radio astronomy or IR detection, \eg, from the James Webb Space Telescope (JWST). 
At the same time, the obtained spectroscopic data in the $1.5\;\upmu$m range allows for
the use of telecommunication grade equipment in the L, S bands to efficiently and 
economically monitor the presence of these two isomers in 
absorption and/ or emission applications.
Finally, the overtone spectrum of \HNCp\ helps to constrain the prediction of the pure 
rotational spectrum of \HNCp, the lower energy isomer of \ce{[CHN]+}, which is, surprisingly, still not available to this day.

\subsection*{Acknowledgments}
This work was supported by the Max Planck Society,
Engineering and Physical Sciences Research Council (Grant No. EP/W03753X/1), and, 
Czech Science Foundation (GACR 22-05935S).
This article is based upon work from COST Action NanoSpace, CA21126, supported by COST (European Cooperation in Science and Technology).
J.S. has been supported by an ERC Advanced Grant (MissIons: 101020583).
The authors gratefully acknowledge the work of the electrical and mechanical workshops and engineering  departments of the Max Planck Institute for Extraterrestrial Physics.
We thank Prof. Stephan Schlemmer (Univ. zu Köln) for lending of the Agilent laser system.


\section*{AUTHOR DECLARATIONS}

\subsection*{Conflict of Interest}
The authors have no conflicts to disclose.

\subsection*{Author Contributions}
Paola Caselli: Conceptualization, Funding acquisition, Investigation, Writing -- review \& editing.
Petr Dohnal: Conceptualization, Funding acquisition, Investigation, Writing -- original draft.
Miguel Jim{\'e}nez-Redondo: Conceptualization, Data curation, Investigation, Writing -- review \& editing.
Pavol Jusko: Conceptualization, Data curation, Investigation, Visualization, Writing -- original draft.

\section*{DATA AVAILABILITY}
The data that support the findings of this study are openly available at
\url{https://doi.org/10.5281/zenodo.15726961}.

\bibliographystyle{elsarticle-num-names}
\bibliography{lit.bib}

\end{document}